\pdfoutput=1
\documentclass[twocolumn]{aastex62}

\usepackage{amsmath}
\usepackage{booktabs}
\usepackage{graphicx}
\usepackage{tabularx}
\usepackage{units}

\usepackage{hyperref}

\newcommand{\Gp}[1] {\ensuremath{\left(#1\right)}}
\newcommand{\Gb}[1] {\ensuremath{\left[#1\right]}}
\newcommand{\GB}[1] {\ensuremath{\left\{#1\right\}}}

\newcommand{\cL}{\ensuremath{\mathcal{L}}}
\newcommand{\cS}{\ensuremath{\mathcal{S}}}

\newcommand{\cB}{\ensuremath{\mathcal{B}}}
\newcommand{\cT}{\ensuremath{\mathcal{T}}}

\begin{document}

\title{Search for Sources of Astrophysical Neutrinos Using Seven Years of
IceCube Cascade Events}
\shortauthors{M.~G.~Aartsen et al.}


\newcommand{\TXS}{TXS~0506+056}

\begin{abstract}
  Low background searches for astrophysical neutrino sources anywhere in the
  sky can be performed using cascade events induced by neutrinos of all flavors
  interacting in IceCube with energies as low as $\sim\unit[1]{TeV}$.
  Previously, we showed that even with just two years of data, the resulting
  sensitivity to sources in the southern sky is competitive with IceCube and
  ANTARES analyses using muon tracks induced by charge current muon neutrino
  interactions | especially if the neutrino emission follows a soft energy
  spectrum or originates from an extended angular region.  Here, we extend that
  work by adding five more years of data, significantly improving the cascade
  angular resolution, and including tests for point-like or diffuse Galactic
  emission to which this dataset is particularly well-suited.  For many of the
  signal candidates considered, this analysis is the most sensitive of any
  experiment.  No significant clustering was observed, and thus many of the
  resulting constraints are the most stringent to date.  In this paper we will
  describe the improvements introduced in this analysis and discuss our results
  in the context of other recent work in neutrino astronomy.
\end{abstract}

\affiliation{III. Physikalisches Institut, RWTH Aachen University, D-52056 Aachen, Germany}
\affiliation{Department of Physics, University of Adelaide, Adelaide, 5005, Australia}
\affiliation{Dept. of Physics and Astronomy, University of Alaska Anchorage, 3211 Providence Dr., Anchorage, AK 99508, USA}
\affiliation{Dept. of Physics, University of Texas at Arlington, 502 Yates St., Science Hall Rm 108, Box 19059, Arlington, TX 76019, USA}
\affiliation{CTSPS, Clark-Atlanta University, Atlanta, GA 30314, USA}
\affiliation{School of Physics and Center for Relativistic Astrophysics, Georgia Institute of Technology, Atlanta, GA 30332, USA}
\affiliation{Dept. of Physics, Southern University, Baton Rouge, LA 70813, USA}
\affiliation{Dept. of Physics, University of California, Berkeley, CA 94720, USA}
\affiliation{Lawrence Berkeley National Laboratory, Berkeley, CA 94720, USA}
\affiliation{Institut f{\"u}r Physik, Humboldt-Universit{\"a}t zu Berlin, D-12489 Berlin, Germany}
\affiliation{Fakult{\"a}t f{\"u}r Physik {\&amp;} Astronomie, Ruhr-Universit{\"a}t Bochum, D-44780 Bochum, Germany}
\affiliation{Universit{\'e} Libre de Bruxelles, Science Faculty CP230, B-1050 Brussels, Belgium}
\affiliation{Vrije Universiteit Brussel (VUB), Dienst ELEM, B-1050 Brussels, Belgium}
\affiliation{Dept. of Physics, Massachusetts Institute of Technology, Cambridge, MA 02139, USA}
\affiliation{Dept. of Physics and Institute for Global Prominent Research, Chiba University, Chiba 263-8522, Japan}
\affiliation{Dept. of Physics and Astronomy, University of Canterbury, Private Bag 4800, Christchurch, New Zealand}
\affiliation{Dept. of Physics, University of Maryland, College Park, MD 20742, USA}
\affiliation{Dept. of Astronomy, Ohio State University, Columbus, OH 43210, USA}
\affiliation{Dept. of Physics and Center for Cosmology and Astro-Particle Physics, Ohio State University, Columbus, OH 43210, USA}
\affiliation{Niels Bohr Institute, University of Copenhagen, DK-2100 Copenhagen, Denmark}
\affiliation{Dept. of Physics, TU Dortmund University, D-44221 Dortmund, Germany}
\affiliation{Dept. of Physics and Astronomy, Michigan State University, East Lansing, MI 48824, USA}
\affiliation{Dept. of Physics, University of Alberta, Edmonton, Alberta, Canada T6G 2E1}
\affiliation{Erlangen Centre for Astroparticle Physics, Friedrich-Alexander-Universit{\"a}t Erlangen-N{\"u}rnberg, D-91058 Erlangen, Germany}
\affiliation{Physik-department, Technische Universit{\"a}t M{\"u}nchen, D-85748 Garching, Germany}
\affiliation{D{\'e}partement de physique nucl{\'e}aire et corpusculaire, Universit{\'e} de Gen{\`e}ve, CH-1211 Gen{\`e}ve, Switzerland}
\affiliation{Dept. of Physics and Astronomy, University of Gent, B-9000 Gent, Belgium}
\affiliation{Dept. of Physics and Astronomy, University of California, Irvine, CA 92697, USA}
\affiliation{Karlsruhe Institute of Technology, Institut f{\"u}r Kernphysik, D-76021 Karlsruhe, Germany}
\affiliation{Dept. of Physics and Astronomy, University of Kansas, Lawrence, KS 66045, USA}
\affiliation{SNOLAB, 1039 Regional Road 24, Creighton Mine 9, Lively, ON, Canada P3Y 1N2}
\affiliation{Department of Physics and Astronomy, UCLA, Los Angeles, CA 90095, USA}
\affiliation{Department of Physics, Mercer University, Macon, GA 31207-0001}
\affiliation{Dept. of Astronomy, University of Wisconsin, Madison, WI 53706, USA}
\affiliation{Dept. of Physics and Wisconsin IceCube Particle Astrophysics Center, University of Wisconsin, Madison, WI 53706, USA}
\affiliation{Institute of Physics, University of Mainz, Staudinger Weg 7, D-55099 Mainz, Germany}
\affiliation{Department of Physics, Marquette University, Milwaukee, WI, 53201, USA}
\affiliation{Institut f{\"u}r Kernphysik, Westf{\"a}lische Wilhelms-Universit{\"a}t M{\"u}nster, D-48149 M{\"u}nster, Germany}
\affiliation{Bartol Research Institute and Dept. of Physics and Astronomy, University of Delaware, Newark, DE 19716, USA}
\affiliation{Dept. of Physics, Yale University, New Haven, CT 06520, USA}
\affiliation{Dept. of Physics, University of Oxford, Parks Road, Oxford OX1 3PU, UK}
\affiliation{Dept. of Physics, Drexel University, 3141 Chestnut Street, Philadelphia, PA 19104, USA}
\affiliation{Physics Department, South Dakota School of Mines and Technology, Rapid City, SD 57701, USA}
\affiliation{Dept. of Physics, University of Wisconsin, River Falls, WI 54022, USA}
\affiliation{Dept. of Physics and Astronomy, University of Rochester, Rochester, NY 14627, USA}
\affiliation{Oskar Klein Centre and Dept. of Physics, Stockholm University, SE-10691 Stockholm, Sweden}
\affiliation{Dept. of Physics and Astronomy, Stony Brook University, Stony Brook, NY 11794-3800, USA}
\affiliation{Dept. of Physics, Sungkyunkwan University, Suwon 16419, Korea}
\affiliation{Dept. of Physics and Astronomy, University of Alabama, Tuscaloosa, AL 35487, USA}
\affiliation{Dept. of Astronomy and Astrophysics, Pennsylvania State University, University Park, PA 16802, USA}
\affiliation{Dept. of Physics, Pennsylvania State University, University Park, PA 16802, USA}
\affiliation{Dept. of Physics and Astronomy, Uppsala University, Box 516, S-75120 Uppsala, Sweden}
\affiliation{Dept. of Physics, University of Wuppertal, D-42119 Wuppertal, Germany}
\affiliation{DESY, D-15738 Zeuthen, Germany}

\author{M. G. Aartsen}
\affiliation{Dept. of Physics and Astronomy, University of Canterbury, Private Bag 4800, Christchurch, New Zealand}
\author{M. Ackermann}
\affiliation{DESY, D-15738 Zeuthen, Germany}
\author{J. Adams}
\affiliation{Dept. of Physics and Astronomy, University of Canterbury, Private Bag 4800, Christchurch, New Zealand}
\author{J. A. Aguilar}
\affiliation{Universit{\'e} Libre de Bruxelles, Science Faculty CP230, B-1050 Brussels, Belgium}
\author{M. Ahlers}
\affiliation{Niels Bohr Institute, University of Copenhagen, DK-2100 Copenhagen, Denmark}
\author{M. Ahrens}
\affiliation{Oskar Klein Centre and Dept. of Physics, Stockholm University, SE-10691 Stockholm, Sweden}
\author{C. Alispach}
\affiliation{D{\'e}partement de physique nucl{\'e}aire et corpusculaire, Universit{\'e} de Gen{\`e}ve, CH-1211 Gen{\`e}ve, Switzerland}
\author{K. Andeen}
\affiliation{Department of Physics, Marquette University, Milwaukee, WI, 53201, USA}
\author{T. Anderson}
\affiliation{Dept. of Physics, Pennsylvania State University, University Park, PA 16802, USA}
\author{I. Ansseau}
\affiliation{Universit{\'e} Libre de Bruxelles, Science Faculty CP230, B-1050 Brussels, Belgium}
% ...
\date{\today}

\collaboration{IceCube Collaboration}
\noaffiliation


\section{Introduction}
\label{sec:intro}

Neutrino astronomy promises to reveal secrets of distant astrophysical objects
that likely can never be observed through other messenger particles.  Because
neutrinos only interact weakly, they can reach us from enormous distances with
no attenuation by intervening matter or background radiation and without
deflection by magnetic fields.  Because they are only produced by hadronic
processes, high energy neutrinos are tracers of high energy cosmic ray
production \citep{Halzen:2002pg}.  While electromagnetic observations can
establish that a source candidate provides sufficient energy density for cosmic
ray acceleration, direct cosmic ray observation is hindered by magnetic
deflection at lower energies and by attenuation at higher energies.  Therefore
neutrino astronomy may offer our best chance for identifying the sources of
high energy cosmic rays \citep{Ahlers:2018fkn}.

Neutrino observation is performed by detecting the Cherenkov radiation emitted
by relativistic charged particles produced when neutrinos collide with matter
in or near a Cherenkov detector.  IceCube, the largest such detector to date,
consists of an array of photomultiplier tubes (PMTs) spanning one $\unit{km^3}$
deep in the Antarctic glacial ice near the geographic South Pole.  IceCube is
sensitive to all neutrino flavors and interaction types.  Charged current (CC)
muon neutrino interactions yield long-lived muons that can travel several
kilometers through the ice \citep{Chirkin:2004hz}, leading to a \emph{track}
signature in the detector.  Neutral current (NC) interactions, and CC
interactions of most other flavors, yield hadronic and electromagnetic showers
that typically range less than \unit[20]{m}~\citep{Aartsen:2013vja}, with 90\%
of the light emitted within \unit[4]{m} of the shower
maximum~\citep{Radel:2012ij}.  The small spatial extent of these showers
compared to the PMT spacing and the scattering length of light in the ice
\citep{Aartsen:2013rt} results in a nearly symmetric \emph{cascade} signature
in the detector.

In 2014, we reported the first observation of a flux of neutrinos above
$\sim\unit[60]{TeV}$ inconsistent with the expectation from atmospheric
backgrounds at greater than $5\sigma$ significance \citep{Aartsen:2014gkd}.
While this measurement was dominated by cascade events, the result was soon
confirmed using muon tracks above $\sim\unit[300]{TeV}$ originating in the
northern sky~\citep{Aartsen:2015rwa,Aartsen:2016xlq}.

More recently, IceCube data revealed the first direct evidence for high energy
neutrino emission associated with a specific astrophysical source, the
gamma-ray blazar \TXS~\citep{IceCube:2018dnn,IceCube:2018cha}.  Before and
since, no other high energy astrophysical neutrino sources have been identified
\citep[e.g.][]{Aartsen:2016oji}.  Most source searches have focused on the muon
track channel, which gives excellent sensitivity to \emph{upgoing} muon tracks
induced by CC muon neutrino interactions.  As viewed by IceCube, upgoing events
correspond to sources in the northern celestial hemisphere.

In much of the southern sky, due to larger background rates, the sensitivity of
the muon track channel to sources following an $E^{-2}$ spectrum is weaker by
an order of magnitude \citep{Aartsen:2016oji} | this factor increases to two
orders of magnitude or more if the spectrum is as soft as $E^{-3}$ or if it has
a cutoff at $E_\text{cut}\lesssim\unit[100]{TeV}$ (see
e.g.~\citet{Aartsen:2017eiu}).

In an initial analysis of two years of data, we demonstrated that the
sensitivity of IceCube in the southern sky can be improved significantly by
performing complimentary searches using cascade events arising from neutrino
interactions of all flavors~\citep{Aartsen:2017eiu}.  Here, we extend that work
in a number of ways.  First, we apply similar, though slightly improved, event
selection criteria to seven years of data.  Second, we obtain significantly
improved cascade angular resolution through the use of a specially-designed
Deep Neural Network.  Finally, we study additional point-like and diffuse
Galactic emission scenarios to which this analysis is expected to be especially
sensitive.  For many of the signal candidates considered, this analysis is the
most sensitive of any experiment to date.  In this paper, we will begin by
describing the IceCube detector and the cascade event selection and
reconstruction.  Then we will introduce the astrophysical neutrino source
candidates considered and the design and performance characteristics of the
statistical methods used.  Finally, we will present our results and discuss
them in the context of other recent work in neutrino astronomy.

\section{IceCube}
\label{sec:icecube}

The IceCube detector (\citet{DETPAPER:1748:0221:12:03:P03012}) is composed of
5160 Digital Optical Modules (DOMs) buried at depths of \unit[1450]{m} to
\unit[2450]{m} in the glacial ice near the geographic South Pole.  Each DOM
includes a 10" photomultiplier tube (PMT) and custom supporting electronics
\citep{Abbasi:2010vc}.  The DOMs are mounted on 86 vertical \emph{strings}
holding 60 DOMs each, arranged in an approximately hexagonal grid.
Seventy-eight of the strings forming the bulk of the array are spaced
\unit[125]{m} apart horizontally, with uniform vertical DOM spacing of
$\unit[\sim17]{m}$.  The remaining 8 strings, which are concentrated near the
center of the detector with $\unit[30-60]{m}$ horizontal spacing, constitute a
denser in-fill array called DeepCore \citep{Collaboration:2011ym}.  On each of
the DeepCore strings, 50 of the DOMs are located in the exceptionally clear ice
at depths of \unit[2100]{m} to \unit[2450]{m}, with vertical spacing of
$\unit[7]{m}$.  The strings were deployed during the Austral summers of
2004--2011.

Digital readouts are triggered when at least eight DOMs observe a signal above
1/4 of the mean expected voltage from a single photoelectron (PE), each in
coincidence with such a signal on a nearest or next-nearest neighboring DOM,
within a $\unit[6.4]{\mu s}$ time window.  When this criterion is met, the data
acquisition system (DAQ, \citet{Abbasi:2008aa}) collects the data from all DOMs
into an \emph{event} and initiates a first round of processing.  Each waveform
is decomposed into series of pulse arrival times and PE counts for use by event
reconstruction algorithms \citep{Ahrens:2003fg,Aartsen:2013vja}.  Simple
selection criteria are applied to reject the most unambiguous cosmic
ray-induced muon backgrounds, reducing the data rate from $\unit[\sim2.7]{kHz}$
at trigger level to $\unit[\sim40]{Hz}$ at filter level.  The filtered dataset
is compressed and transmitted via satellite to a data center in the north for
further processing.

\section{Dataset}
\label{sec:selreco}

After the initial selection applied at the South Pole, the remaining dataset is
still dominated by atmospheric muons.  In order to search for neutrino sources,
neutrino candidates are selected, and their properties are reconstructed based
on the light arrival pattern observed in the DOMs.  In the following, we
discuss a re-optimized method for selecting neutrino-induced cascades and a
novel machine learning-based approach to reconstructing their arrival
directions and energies.

\subsection{Event Selection}
\label{sec:selreco:sel}

The procedure for rejecting the atmospheric muon background depends on the
event topology of interest.  Neutrino-induced muon tracks with energies
$\gtrsim\unit[1]{TeV}$ originating in the northern sky can be selected with
high efficiency and low atmospheric muon contamination by identifying events
reconstructed at declinations $\delta\gtrsim5^\circ$ with high confidence,
as only neutrinos can travel through so much intervening earth and/or ice
before producing muons that pass through the detector.  Neutrino- and cosmic
ray-induced muon tracks originating in the southern sky and entering the
detector from above can only be distinguished probabilistically, and only
under the assumption that the neutrino spectrum is harder than the
atmospheric muon spectrum.  Thus the energy threshold increases to
$\sim\unit[100]{TeV}$ in the southern sky, resulting in weaker sensitivity
especially for a soft neutrino spectrum.

In this work, we instead turn our attention to cascade events produced when the
neutrino interaction vertex, and hence first observed light, occurs inside the
detector.  With this approach we accept all neutrino flavors and most
interaction types, approximately independent of declination, while efficiently
rejecting downgoing atmospheric muons.  An added benefit for astrophysical
neutrino searches is that for declinations $\lesssim-30^\circ$ the atmospheric
neutrino background is naturally suppressed because many are accompanied by
incoming atmospheric muons originally produced in the same cosmic ray shower in
the upper atmosphere \citep{Schonert:2008is}.

Most Cherenkov light from a muon traveling through ice is radiated through
stochastic processes, resulting in a dense, linear series of cascade-like
signatures that may be observed in our detector.  The mean distance between
these energy deposits decreases with increasing energy.  For energies
$\gtrsim\unit[60]{TeV}$, incoming muons can be rejected with high confidence
using a veto region consisting of just the outermost DOMs, reserving the
majority of the instrumented region as a fiducial volume for neutrino detection
\citep{Aartsen:2014gkd}.  To lower the threshold to $\sim\unit[1]{TeV}$ while
holding the incoming muon rejection rate constant, the thickness of the veto
region must be increased.  Below we summarize this method, which is used as
described in \citet{Weaver:2017ypp} and which further optimizes the approach
first introduced in \citet{Aartsen:2014muf}.

We begin with all events passing one or more basic filters at the South Pole.
A splitting algorithm is applied to each event, identifying $\sim75\%$ of
unrelated but temporally coincident physical events initially merged in the DAQ
output by clustering causally connected sets of pulses.  We reject any event in
which the first $\ge3$ pulses appear in the outer layer veto region as
described in \cite{Aartsen:2013jdh}.  An additional veto is applied to reject
events in which two or more PE are observed consistent with a downgoing track
passing through the interaction vertex or a major energy deposition.  Finally,
a cut is applied on the interaction vertex location, scaling with observed
charge as described in \cite{Aartsen:2014muf} such that at \unit[100]{PE} the
fiducial volume is reduced to just the DeepCore sub-array, while at
$\ge\unit[6000]{PE}$ the fiducial volume consists of all but the outermost
layer of DOMs.  This final cut enables efficient background rejection down to
$\sim\unit[1]{TeV}$ by keeping the probability of observing veto photons
approximately independent of energy.

We rely on a traditional maximum likelihood method \citep{Aartsen:2013vja} to
obtain initial reconstructions used for cascade/track discrimination.  The goal
of this reconstruction is to unfold the spatial and temporal pattern of energy
depositions for each event.  Two fits are performed: one which is constrained
to find a single dominant cascade-like energy deposition, and one which finds a
linear combination of such energy depositions distributed along a possible muon
track.  Events in which at least $\unit[6000]{PE}$ were collected are
classified as tracks if the free track fit finds at least two non-negligible
depositions more than \unit[500]{m} apart, or if the free track fit is
associated with more charge than the single cascade fit.  Events with less
total collected light are classified as tracks if at least \unit[1.5]{PE}  are
consistent with an outgoing muon track originating at the reconstructed
interaction vertex \citep{Weaver:2017ypp}.  All other events are classified as
cascades and are used in the present analysis.

The selection criteria described above were applied to data taken from May 2010
to May 2017 as well as to neutrino and atmospheric muon Monte Carlo (MC)
simulations used for performance estimates.  The first year of data comes from
the nearly-complete 79-string configuration while the remaining six years make
use of the complete 86-string detector.  In a total of 2428 days of IceCube
livetime, 10422 events survive until cascade/track discrimination; of these,
1980 are identified as cascades.  Note that while the dominant improvement in
this dataset is the increase from two to seven years of data, the neutrino
effective area is also enhanced by applying coincident event splitting and veto
criteria to data from every initial South Pole filter.  This increases the
acceptance by 23\% (67\%) for a signal following an $E^{-2}$ ($E^{-3}$)
spectrum.

From MC simulations, we find that 98\% of truly cascade-like events which pass
all selection criteria are correctly identified as such.  The rate at which CC
muon neutrino interactions are successfully classified as track events
increases with energy as more light is produced by the outgoing muon.  For a
conventional atmospheric neutrino spectrum, 30\% of the cascade channel
consists of misclassified CC muon neutrino interactions; for an astrophysical
spectrum following $E^{-2.5}$ or harder, this contribution reduces to 5\% or
less.  This population of misclassified events results in a tolerable
background at lower energies as well as a small signal contribution at higher
energies.

Because muon track analyses specifically target events with high quality track
reconstructions and reject events dominated by individual cascade-like energy
depositions, we expect the cascade analysis to be largely statistically
independent in spite of the small but nonzero misclassification rate.  In fact,
the final cascade selection shares just a single $\sim\unit[2]{TeV}$ event in
common with the latest muon track selection.

\subsection{Event Reconstruction}
\label{sec:selreco:reco}

In past work, we have used a maximum likelihood method to reconstruct neutrino
energy and direction of travel from IceCube cascades \citep{Aartsen:2013vja}.
This approach relies on detailed parameterizations of the position- and
direction-dependent light absorption and scattering lengths in the ice, neither
of which is large compared to the DOM spacing.  This results in a complex
multi-dimensional likelihood function with many local optima in the right
ascension and declination coordinates $(\alpha,\delta)$, such that it is
computationally expensive to find the global optimum for any given event and
prohibitive to estimate the per-event statistical uncertainties.

In this work, we introduce a novel cascade reconstruction using a deep Neural
Network (NN).  A NN is a highly flexible function mapping from an \emph{input
layer} to an \emph{output layer} via a series of \emph{hidden layers}, where
each successive layer consists of a set of values computed based on the values
contained in the previous layer.  The functional forms of the layer-to-layer
connections (the network \emph{architecture}) must be designed \emph{a priori};
the numerical parameters of those connections are optimized through a training
procedure to yield good results for a given training dataset.  NNs are
well-suited to problems in high energy physics for which we are typically able
to generate high-statistics MC datasets for use in training.

Our NN-based reconstruction draws from recent advances in image recognition
and is implemented using Tensorflow \citep{tensorflow2015:whitepaper}.  The
network architecture used here is largely the same as one introduced previously
for muon energy reconstruction \citep{Huennefeld:2017pdh}.  The method will be
described in detail in a separate publication, but here we will outline the
main considerations relevant in this analysis.

IceCube data consists of a set of waveforms (represented as a series of pulse
arrival times and PE counts) accumulated over time on a number of DOMs
distributed throughout the three-dimensional instrumented volume, and thus is
in general four-dimensional.  Our first step is to compute waveform summary
values for use in the input layer.  For each DOM, these values consist of the
relative time of the first pulse; the time elapsed until 20\%, 50\%, and 100\%
of the total charge is collected; the total charge collected; the charge
collected within \unit[100]{ns} and \unit[500]{ns} of the first pulse; and the
charge-weighted mean and standard deviation of relative pulse arrival times.

The detector is divided into three sub-arrays: IceCube, lower DeepCore, and
upper DeepCore.  Each sub-array is independently well-approximated by a regular
spatial grid suitable for processing by several initial convolutional layers,
which are able to exploit symmetries in the structure of the input data to
facilitate efficient network optimization and usage\footnote{Alternative
methods are being developed to avoid the reliance on regular detector
geometry.} (see \citet{Huennefeld:2017pdh} for diagrams of the relevant
geometry).  The output from the convolutional layers is taken as the input for
each of two fully-connected neural networks (in which each node in a given
layer is connected to every node in the preceding layer).  One of these
networks is optimized to estimate the physical parameters of interest | the
right ascension, declination, and energy $(\alpha,\delta,E)$ | while the other
is optimized to estimate the uncertainties on these parameters.

All training was performed using 50\% of the relevant signal MC, with the
remaining 50\% reserved for testing analysis-level performance.  Two training
passes were performed.  The first pass made use of several MC datasets: one
with baseline values for key parameters such as DOM quantum efficiency and
light absorption and scattering lengths, and several more with modified values
within estimated systematic uncertainties.  In addition to offering overall
increased training statistics, the use of these differing datasets may give the
NN some robustness against known systematic uncertainties.  The second
training pass refined the network to give the smallest errors and, on-average,
unbiased reconstructions for the baseline MC.  In each pass, \emph{a priori}
per-parameter weighting was applied such that angular resolution is valued over
energy resolution by a factor of~5.

\begin{figure}[t]
  \centering
  \includegraphics[width=\columnwidth]{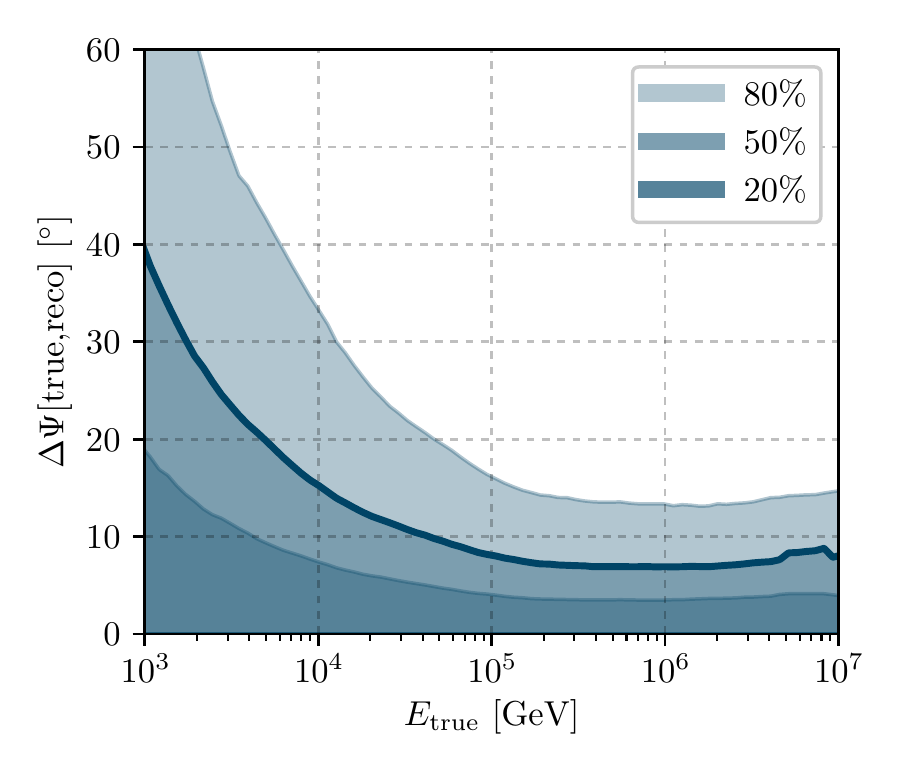}
  \caption{
    Expected angular reconstruction performance as a function of neutrino
    energy, estimated using MC and including systematic uncertainties (see
    Section~\ref{sec:methsens:sys}).  Shaded regions indicate the radii of
    error circles covering 20\%, 50\%, and 80\% of events.
  }
  \label{fig:selrec:angres}
\end{figure}

\begin{figure*}[t]
  \centering
  \includegraphics[width=.49\textwidth]{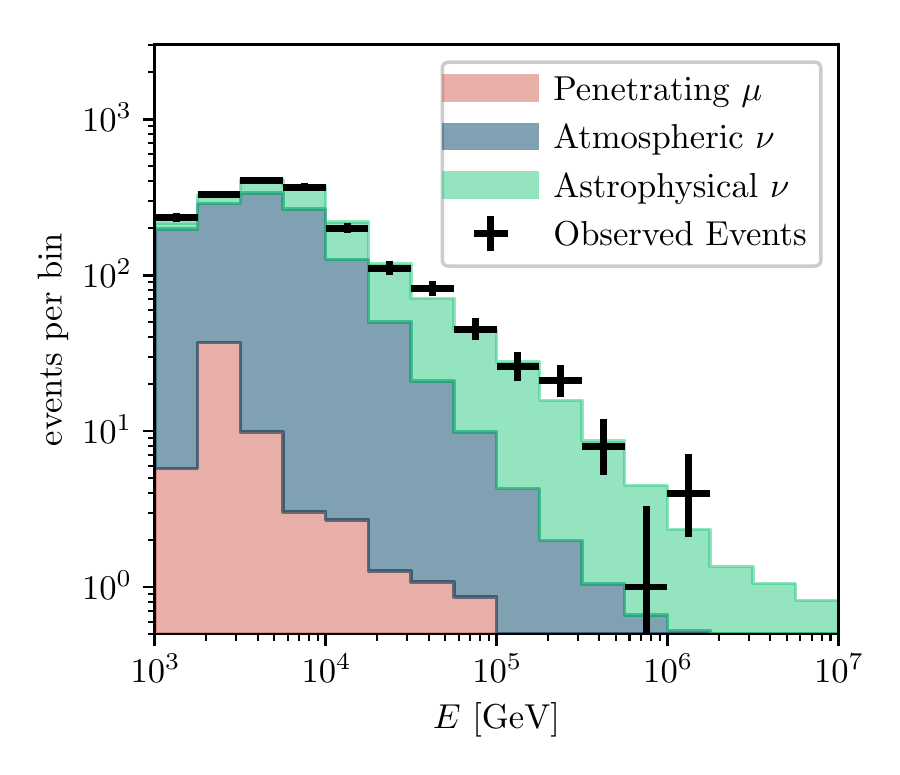}
  \hfill
  \includegraphics[width=.49\textwidth]{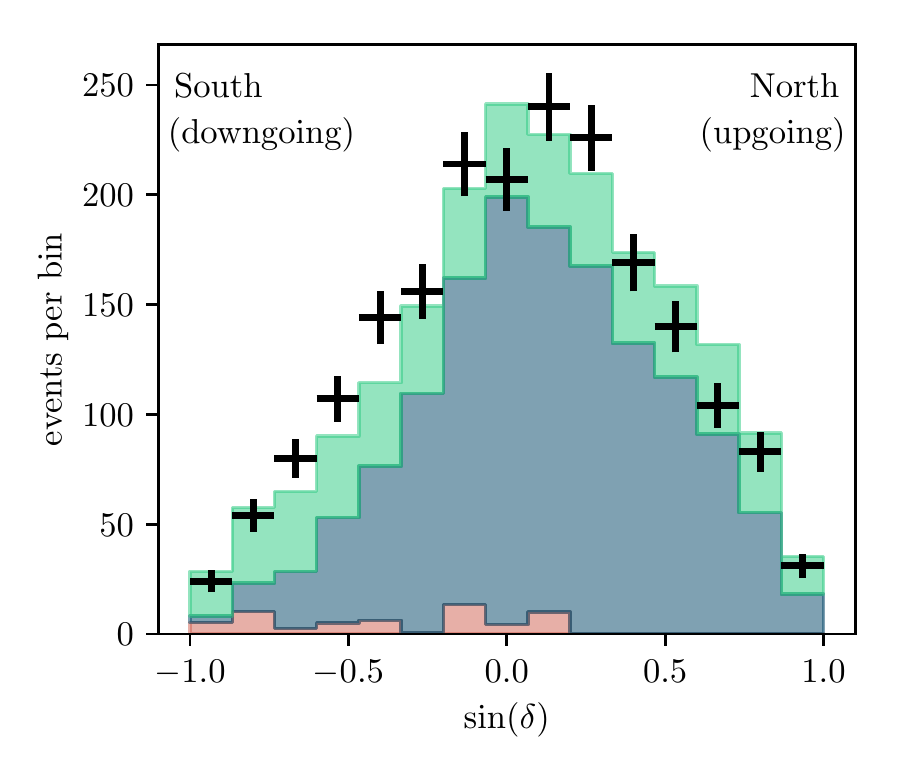}
  \caption{
    Energy and $\sin(\delta)$ distributions for data and MC.  Atmospheric
    muons appear preferentially in the downgoing region, $\sin(\delta)<0$,
    and at energies below \unit[100]{TeV}.  A clear excess of high energy
    events is attributed to astrophysical neutrinos.
  }
  \label{fig:selrec:Esindec}
\end{figure*}

The expected performance of the NN angular reconstruction (including
systematics; see Section~\ref{sec:methsens:sys}) is shown as a function of
energy in Figure~\ref{fig:selrec:angres}.  Compared to the reconstructions used
in our previous analysis of two years of data \citep{Aartsen:2017eiu}, the NN
offers significantly improved angular resolution above \unit[10]{TeV} (a factor
of 2 improvement at \unit[1]{PeV}).  While we do not recover the optimal
statistics-limited angular resolution described in \citet{Aartsen:2013vja}, we
do obtain performance that improves monotonically with increasing energy up to
$\sim\unit[1]{PeV}$.  At higher energies, the estimated systematic uncertainty
becomes large enough to prevent any further improvement.  Note that an
additional advantage of the NN angular reconstruction used here is that it
naturally provides per-event uncertainty estimates usable in the statistical
analysis described in Section~\ref{sec:methsens:meth}, whereas previous work
relied on a parameterization of typical uncertainties derived from signal MC.

The performance of the energy reconstruction is comparable to that used in
previous work.  The estimated energy is within 60\% of the true neutrino energy
for 68\% of events, averaged over all neutrino flavors and interaction types,
and approximately independent of spectrum.  This performance estimate, like the
sensitivities quoted in Section~\ref{sec:methsens:sens}, assumes a flavor ratio
of 1:1:1 with equal contributions from $\nu$ and $\bar\nu$, detected via a
mixture of CC and NC interactions.

The energy and declination distributions of cascade events in data are compared
with neutrino and atmospheric muon MC in Figure~\ref{fig:selrec:Esindec}.  The
distributions obtained are similar to those observed in the two year sample
\citep{Aartsen:2014muf,Aartsen:2017eiu}.

\section{Source Candidates}
\label{sec:cand}

In this work, we search for neutrino emission from a number of Galactic and
extra-Galactic source candidates.  Each candidate has been studied previously
by IceCube, by ANTARES, a neutrino observatory located deep in the
Mediterranean sea \citep{2011NIMPA.656...11A}, or by both, such that direct
comparisons can be drawn between the results presented here and past work using
IceCube tracks and all interaction flavors in ANTARES.  In this section, we
outline the neutrino emission scenarios that we have considered.

\subsection{Point-like Source Candidates}
\label{sec:cand:pt}

One way to search for astrophysical neutrino sources with only a minimal set of
\emph{a priori} assumptions about source position is to search the entire sky
for the most significant point-like neutrino clustering in excess of the
background expectation on a dense grid of pixels that are small compared to the
neutrino angular resolution.  This approach has most recently been employed by
IceCube using tracks \citep{Aartsen:2016oji} and cascades
\citep{Aartsen:2017eiu} as well as by ANTARES using tracks and cascades in
combination \citep{Albert:2017ohr}, and we include it in the present analysis
as well.  However, an all-sky scan is subject to a large trial factor and thus
is in general less sensitive compared to analyses that use prior information to
restrict the set of hypothesis tests.

An alternative approach is to scan only the positions of a modest number of
well-motivated source candidates, which substantially reduces the trial factor.
In addition, where multiple analyses report results for the same or overlapping
catalogs, direct comparisons can be made.  Here we scan the same catalog of 74
source candidates that was studied in the previous IceCube cascade
paper~\citep{Aartsen:2017eiu}.

We consider one source in more detail: the supermassive black hole at the
center of the Galaxy, Sagitarius A*.  Based on hints from gamma-ray
observations \citep[e.g.][]{Herold:2019pei}, there may be emission up to some
unknown high energy cutoff from a spatially extended region centered
approximately on this object.  Therefore we evaluate constraints on the flux
from this region as a function of possible spatial extension and for several
possible spectral cutoffs.

The gamma-ray blazar \TXS\ does not appear in the \emph{a priori} catalog
described above.  In light of this, and in anticipation of future
identifications of unexpectedly promising source candidates based on neutrino
observations, we treat this object as a \emph{monitored source} to be studied
separately from the catalog scan described above.

For source classes for which we can predict approximate relative signal
strengths, it may be possible to increase the signal-to-background ratio using
a source-stacking method \citep[e.g.][]{Abbasi:2011}.  Because the present
analysis offers good sensitivity in the southern sky, roughly independent of
possible spatial extension up to a few degrees, we include stacking analyses
for three Galactic supernova remnant (SNR) catalogs derived from SNR Cat
\citep{2012AdSpR..49.1313F} and previously studied using IceCube tracks
\citep{Aartsen:2017ujz}.  These SNRs are categorized based on their
environment: those with associated molecular clouds, those with associated
pulsar wind nebulae (PWN), and those with neither.  The angular extension of
these objects reach up to $1.63^\circ$, and each catalog comprises a
preponderance of objects in the southern sky.

\subsection{Diffuse Galactic Emission}
\label{sec:cand:gp}

\newcommand{\FermiL}{\emph{Fermi}-LAT}
\newcommand{\Fermi}{\emph{Fermi}}

Cosmic ray interactions with interstellar gas in the Milky Way are expected to
produce neutral and charged pions, where neutral pions would decay to
observable gamma rays and charged pions would yield potentially observable
neutrinos.  The hadronic gamma-ray emission up to \unit[100]{GeV} has been
identified by \FermiL\ using a multi-component fit \citep{Fermi:2012:diffgal}.
A corresponding neutrino flux prediction can be obtained by extrapolating this
measurement to energies above $\unit[1]{TeV}$ in the context of Galactic cosmic
ray production and propagation models.

\newcommand{\KRAga}{\ensuremath{\text{KRA}_\gamma}}
\newcommand{\KRAg}{\ensuremath{\text{KRA}_\gamma^5}}
\newcommand{\KRAG}{\ensuremath{\text{KRA}_\gamma^{50}}}

The original model fits by \emph{Fermi} somewhat under-predict the measured
gamma ray flux in the Galactic plane, and especially near the Galactic center,
above $\unit[10]{GeV}$.  The \KRAga\ models obtain better agreement with gamma
ray data in this regime by introducing galactocentric cosmic ray diffusion
parameter variability and an advective
wind~\citep{Gaggero:2015xza,Gaggero:2017jts}.  Model-dependent neutrino flux
predictions are provided assuming cosmic ray injection spectra with exponential
cutoffs at \unit[5]{PeV} or \unit[50]{PeV} per nucleon; we refer to these as
\KRAg\ and \KRAG, respectively.

The latest constraints on diffuse Galactic neutrino emission depend on the
\KRAga\ models and were obtained in a joint IceCube and ANTARES analysis
\citep{Albert:2018vxw} which made use of complimentary features of the IceCube
track analysis \citep{Aartsen:2017ujz} and ANTARES track and cascade combined
analysis \citep{Albert:2017oba}.  In this work we search for emission following
\KRAg\ as the primary diffuse Galactic emission result; we also test for
emission following \KRAG.  Finally, we test for emission following the spatial
profile of the \FermiL\ $\pi^0$-decay measurement, assuming an $E^{-2.5}$
neutrino energy spectrum.

\subsection{\emph{Fermi} Bubbles}
\label{sec:cand:fb}

The \Fermi\ bubbles consist of a pair of gamma ray emission regions that extend
to $\sim55^\circ$ above and below the Galactic center
\citep{2010ApJ...724.1044S}.  Most of the \Fermi\ bubble region yields a
relatively hard gamma ray spectrum up to $\sim\unit[100]{GeV}$, with some
evidence for spectral softening above that energy \citep{Fermi-LAT:2014sfa}.
The gamma-ray emission has been speculated to be of hadronic origin
\citep{2011PhRvL.106j1102C}, powered by cosmic ray acceleration in the vicinity
of the Galactic Center; however, the true origin of the \Fermi\ bubbles has not
yet been experimentally identified.

We derive constraints on emission from the \Fermi\ bubbles following spectra of
the form $dN/dE\propto E^{-2.18}\cdot\exp(-E/E_\text{cut})$, for 
$E_\text{cut}\in\GB{\unit[50]{TeV}, \unit[100]{TeV}, \unit[500]{TeV}}$ | the
same spectra tested in recent work by ANTARES \citep{Hallmann:2017mac}.  If there is
neutrino emission from the \Fermi\ bubbles with a significantly softer spectrum
or lower cutoff energy, this analysis would not be sensitive to it.


\section{Analysis Methods and Performance}
\label{sec:methsens}

The source searches described in the previous section use established
methods from recent IceCube work.  In this section, we review the
statistical methods and describe the systematic uncertainty treatment
applied here.  Then we discuss the sensitivity of this analysis to the
source candidates under consideration.

\subsection{Statistical Methods}
\label{sec:methsens:meth}

In this work we consider two broad categories of source candidates: point-like
and extended template, where the latter include diffuse Galactic emission and
emission spanning the \Fermi\ Bubbles.  Both analysis types are based on the
standard likelihood \citep{Braun:2008bg} given by a product over all events $i$
in the dataset:
\begin{gather}
  \cL(n_s,\gamma)
  = \prod_i \Gb{
    \frac{n_s}{N}\cS_i(\vec{x}_i|\gamma)
    + \Gp{1 - \frac{n_s}{N}}\cB_i(\vec{x}_i)
  },
  \label{eq:L}
\end{gather}
where $N$ is the total number of events; $n_s$ is the expected number of signal
events; $\gamma$ is the signal spectral index; $\vec{x}_i$ represents the event
right ascension, declination, angular uncertainty, and energy
$\GB{\alpha_i,\delta_i,\sigma_i,E_i}$; $\cS_i(\vec{x}_i|\gamma)$ is the
probability density function (PDF) assuming event $i$ is part of the signal
population; and $\cB_i(\vec{x}_i)$ is the PDF assuming event $i$ is part of the
atmospheric or unrelated astrophysical background populations.  For all source
types, $n_s$ is free to vary between 0 and $N$.  For point-like sources, the
signal spectral index $\gamma$ is free to vary between 1 and 4, while for
extended templates $\gamma$ is fixed to a source-dependent constant value
($\gamma=2.5$ for diffuse Galactic emission and $\gamma=2.18$ for emission
from the \Fermi\ bubbles).

The details of our signal and background likelihoods, $\cS_i$ and $\cB_i$,
follow established methods applied previously to IceCube tracks for
individual \citep{Aartsen:2016oji} and stacked \citep[e.g.][]{Abbasi:2011}
point-like sources as well as for extended templates
\citep{Aartsen:2017ujz}.  We do not require a specialized treatment, in
contrast to our previous cascade analysis \citep{Aartsen:2017eiu}, thanks to
increased statistics in the experimental dataset as well as new per-event
angular uncertainty estimates given by the NN reconstruction.

As in previous work, we define the test statistic as the log likelihood ratio
$\cT=-2\ln\{\cL(n_s=0)/\cL(\hat{n}_s,\hat{\gamma})\}$, where $\hat{n}_s$ and
$\hat\gamma$ are the values which maximize $\cL$, subject to the constraints
specified above.  This test statistic is used to compute significances,
sensitivities, discovery potentials, and upper limits (ULs).  For the all-sky
(source candidate catalog) scan, we compute a post-trials significance based on
the most significant pixel (source candidate) tested, in order to guarantee the
reported false positive rates.  Sensitivities (90\% CL), upper limits (90\%
CL), and discovery potentials ($5\sigma$) are defined as in our previous
analysis \citep{Aartsen:2017eiu} and are computed using the Neyman construction
\citep{neyman}.

\subsection{Systematic Uncertainties}
\label{sec:methsens:sys}

The dominant systematic uncertainties in this analysis include the optical
properties of the ice, the quantum efficiency of the DOMs, and the neutrino
interaction cross section.  These uncertainties affect the angular resolution
and the signal acceptance.  As in our previous cascade analysis
\citep{Aartsen:2017eiu}, we treat these effects as approximately separable.
However, we have improved our approach to each consideration; we describe our
latest method in the following.

The NN reconstruction is trained to yield optimal performance on baseline MC;
the angular resolution for real data events will be somewhat worse.  To
estimate how much worse, we perform dedicated simulations of events similar to
those observed, but using depth-dependent ice model variations intended to
cover the uncertainties in the model.  By comparing the median resolution from
these modified simulations with that from the baseline MC, we obtain a function
of energy that quantifies how much worse the resolution may be than expected
from the baseline.  This factor ranges from 10\% at \unit[1]{TeV} to $\sim50\%$
at \unit[2]{PeV}, and is taken as a correction to the angular separation
between the reconstructed and true direction for each event in the baseline MC.
This factor is similarly applied to the angular uncertainty estimates
$\sigma_i$ for both MC and data events.  In this way, we directly account for
systematic uncertainties impacting angular resolution in the quantiles shown
in Figure~\ref{fig:selrec:angres} as well as in all p-values and sensitivity
flux calculations in the analysis.

The above treatment accounts for the analysis-level impact of systematic
uncertainties for each observed event.  To address the uncertainties in the
detection efficiency, and thus in sensitivity, discovery potential, and upper
limit fluxes, we compute the energy-integrated signal acceptance, as a function
of declination and for each considered spectrum, based on additional MC
datasets produced with varied modeling assumptions (the same modified datasets
used in NN training; see Section~\ref{sec:selreco:reco}).  We find that for
plausible ice model and detector variations, the signal acceptance variation
ranges from $\sim10\%$ for an unbroken $E^{-2}$ spectrum to $\sim17\%$ for
$E^{-2}$ with an exponential cutoff at \unit[100]{TeV}, roughly independent of
declination.  As was done in the previous analysis, we estimate an uncorrelated
$4\%$ impact from uncertainties in the neutrino interaction cross section.
These values are added in quadrature on a per-spectrum basis to obtain a final
estimate of uncertainties via signal acceptance effects.  In the remainder of
this paper, all sensitivity, discovery potential, and upper limit fluxes
include this factor.

\begin{figure*}[ht]
  \centering
  \includegraphics[width=\textwidth]{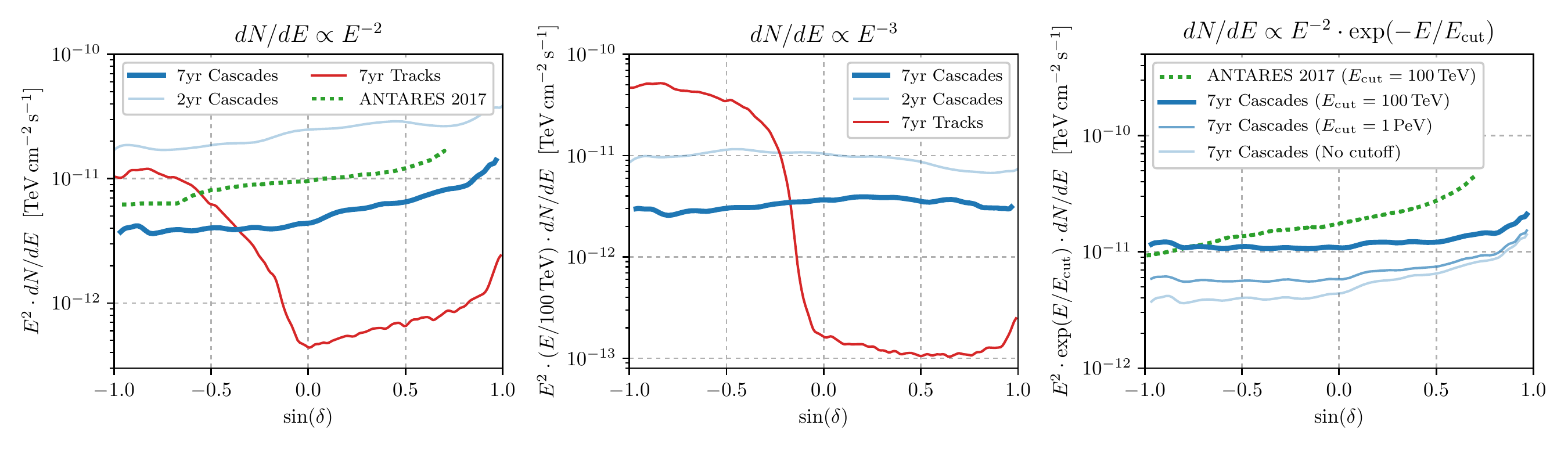}
  \caption{
    Per-flavor sensitivity as a function of $\sin(\delta)$ to point sources
    following an unbroken $E^{-2}$ spectrum (left), unbroken $E^{-3}$ spectrum
    (center), and $E^{-2}$ spectrum with some possible exponential cutoffs
    (right).  This work is labeled as \emph{7yr Cascades}.  Past IceCube work
    shown here includes includes \emph{2yr Cascades} \citep{Aartsen:2017eiu}
    and \emph{7yr Tracks} \citep{Aartsen:2016oji}; ANTARES curves are taken
    from \cite{Albert:2017ohr}.
  }
  \label{fig:methsens:senswide}
\end{figure*}

\subsection{Sensitivity}
\label{sec:methsens:sens}

All sensitivities discussed in the remainder of this paper are per-neutrino
flavor (assuming a flavor ratio of 1:1:1 at the detector), but summed over
$\nu$ and~$\bar\nu$.  The point source sensitivity flux as a function of source
declination is shown in Figure~\ref{fig:methsens:senswide} for several spectral
scenarios: unbroken power laws following hard ($\gamma=2$) and soft
($\gamma=3$) spectra, and spectral cutoff scenarios $dN/dE\propto
E^{-2}\cdot\exp(-E/E_\text{cut})$ with
$E_\text{cut}\in(\unit[100]{TeV},\,\unit[1]{PeV},\,\gg\unit[1]{PeV})$.  Where
published values are available for previous IceCube work with tracks
\citep{Aartsen:2016oji} or cascades \citep{Aartsen:2017eiu}, or for the most
recent ANTARES track and cascade combined analysis \citep{Albert:2017ohr},
these are shown for comparison.  We find that the present analysis improves
upon the previous IceCube work with cascades at all declinations and across the
tested spectra, with the largest improvements reaching a factor larger than 4
in the southern sky.  Furthermore, we now obtain the best sensitivity of any
analysis for hard sources in the southern-most $\sim30\%$ of the sky
($\sin(\delta)<-0.4$).  This search also achieves sensitivity comparable to
that of ANTARES for spectra with cutoffs as low as
$E_\text{cut}=\unit[100]{TeV}$, but with much weaker declination dependence.

The sensitivities of the SNR stacking analyses are listed in
Table~\ref{tab:sensres:snr}.  In this work we obtain a sensitivity below
previously set ULs \citep{Aartsen:2017ujz} only for the SNR-with-PWN catalog,
which consists of eight southern SNRs and one northern SNR.  It is nevertheless
interesting to revisit all three catalogs here because, while they all include
southern source candidates, in previous work the results necessarily were
dominated by northern candidates due to the strongly declination-dependent
signal acceptance of the IceCube track selection.

The sensitivities of the diffuse Galactic template analyses are listed in
Table.~\ref{tab:sensres:gp}.  This analysis obtains \mbox{$\sim30\%\ (40\%)$}
better sensitivity to \KRAg\ (\KRAG) than the recent joint IceCube+ANTARES
analysis \citep{Albert:2018vxw}.  Compared to the IceCube analysis using seven
years of tracks \citep{Aartsen:2017ujz}, this analysis obtains $\sim15\%$
better sensitivity to emission following the spatial profile of the \FermiL\
$\pi^0$-decay measurement.  These improvements are possible because the
expected emission follows a soft ($\gamma\sim2.5$) spectrum and is concentrated
near the Galactic center at $\delta\sim-30^\circ$, where IceCube track analyses
are subject to a large background of atmospheric muons but the present cascade
analysis efficiently rejects this background as well as some of the atmospheric
neutrino background; the improvement is larger for the \KRAga\ models than for
the \FermiL\ $\pi^0$ model because the former are specifically tuned to
increase the concentration of the expected flux near the Galactic center.

The sensitivity flux for the \Fermi\ Bubble analyses is $\sim30\%$ below the
upper limits shown in Figure~\ref{fig:concout:fb}, approximately independent of
spectral cutoff.  This analysis obtains sensitivity that is at least one order
of magnitude better than the recent ANTARES search \citep{Hallmann:2017mac}, with
the improvement increasing with spectral cutoff energy, $E_\text{cut}$.
Because we assume an even more extended template than ANTARES, covering a total
solid angle of about \unit[1.18]{sr} compared to $\sim\unit[0.66]{sr}$, this
factor is even larger if considered in terms of flux per solid angle.  Once
again, this improvement is due to efficient rejection of atmospheric
backgrounds for the cascade dataset used in this work.

\begin{figure*}[ht]
  \centering
  \includegraphics[width=\textwidth]{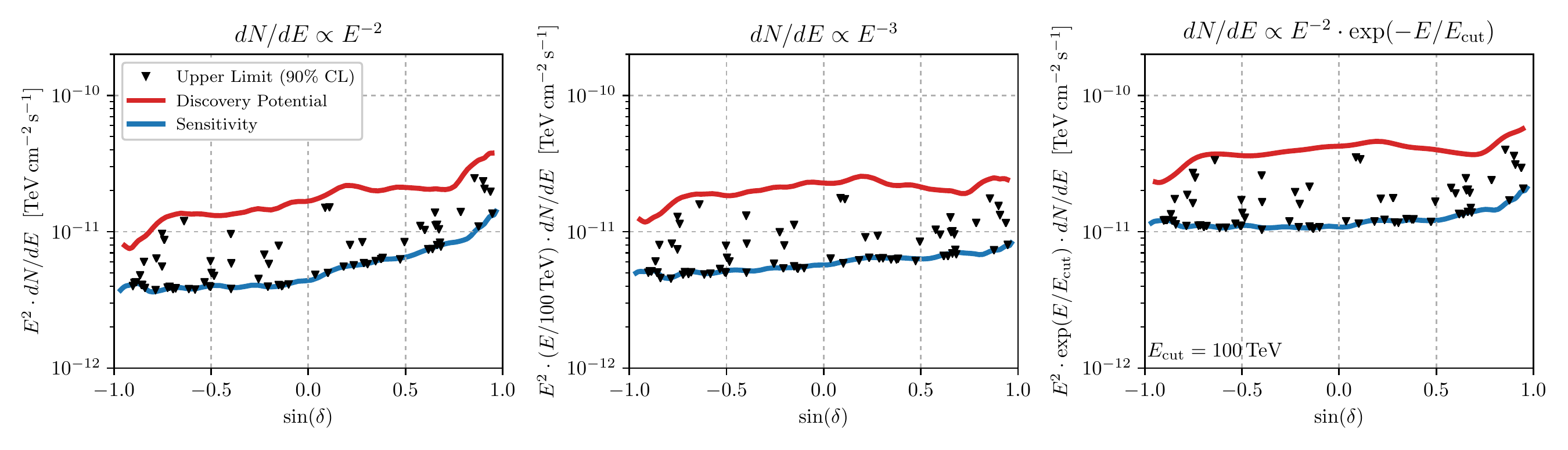}
  \caption{
    Per-flavor sensitivity, discovery potential, and source candidate upper
    limits as a function of $\sin(\delta)$, for point sources following an
    unbroken $E^{-2}$ spectrum (left), unbroken $E^{-3}$ spectrum (center),
    and $E^{-2}$ spectrum with an exponential cutoff at
    $E_\text{cut}=\unit[100]{TeV}$ (right).
  }
  \label{fig:results:sensdisclim}
\end{figure*}

\section{Results}
\label{sec:results}

\begin{table*}
  \centering
  \begin{tabular}{l c c c c c c c c c}
    \toprule
    & \multicolumn{5}{c}{7yr Cascades}
    & \multicolumn{4}{c}{7yr Tracks} \\
    \cmidrule(lr){2-6} \cmidrule(lr){7-10}
    Catalog
    &Sensitivity & p-value & $n_s$ & $\gamma$ & UL
    & p-value & $n_s$ & $\gamma$ & UL \\
    \midrule
    SNR with mol.~cloud
    & 9.9 & 0.12 & 17.2 & 3.76 & 24
    & 0.25 & 16.5 & 3.95 & 2.23 \\
    SNR with PWN
    & 6.3 & 1 & 0 & | & 6.3 
    & 0.34 & 9.36 &  3.95 & 11.7 \\
    SNR alone
    & 7.5 & 0.082 & 8.2 & 2.42 & 15
    & 0.42 & 3.82 & 2.25 & 2.06 \\
    \bottomrule
  \end{tabular}
  \caption{
    Sensitivity and results of the SNR stacking analyses, compared to the
    previous analysis with tracks \citep{Aartsen:2017ujz}.  Sensitivity and
    ULs are given as $E^2\cdot (E/\unit[100]{TeV})^{0.5}\cdot dN/dE$ in
    units $\unit[10^{-12}]{TeV\,cm^{-2}\,s}$.
  }
  \label{tab:sensres:snr}
\end{table*}

\begin{table*}
  \centering
  \begin{tabular}{l c c c c c c c c}
    \toprule
    & \multicolumn{4}{c}{7yr Cascades}
    & \multicolumn{4}{c}{Previous Work} \\
    \cmidrule(lr){2-5} \cmidrule(lr){6-9}
    Template
    & p-value & Sensitivity & Fitted Flux & UL
    & p-value & Sensitivity & Fitted Flux & UL \\
    \midrule
    \KRAg
    & 0.021 & 0.58 & 0.85 & 1.7
    & 0.29 & 0.81 & 0.47 & 1.19 \\
    \KRAG
    & 0.022 & 0.35 & 0.65 & 0.97
    & 0.26 & 0.57 & 0.37 & 0.90 \\
    \FermiL\ $\pi^0$
    & 0.030 & 2.5 & 3.3 & 6.6
    & 0.37 & 2.97 & 1.28 & 3.83 \\
    \bottomrule
  \end{tabular}
  \caption{
    Sensitivity and results of the diffuse Galactic template analyses,
    compared to latest previous work: a joint IceCube-ANTARES
    \citep{Albert:2018vxw} for \KRAga\ models, and seven years of IceCube
    tracks \citep{Aartsen:2017ujz} for \FermiL $\pi^0$ decay.  Sensitivity,
    fitted flux, and ULs are given as multiples of the model prediction for
    \KRAga\ models, and as $E^2\cdot (E/\unit[100]{TeV})^{0.5}\cdot dN/dE$
    in units $\unit[10^{-11}]{TeV\,cm^{-2}\,s^{-1}}$ for \FermiL\ $\pi^0$
    decay.
  }
  \label{tab:sensres:gp}
\end{table*}

\begin{figure}[t]
  \includegraphics[width=\columnwidth]{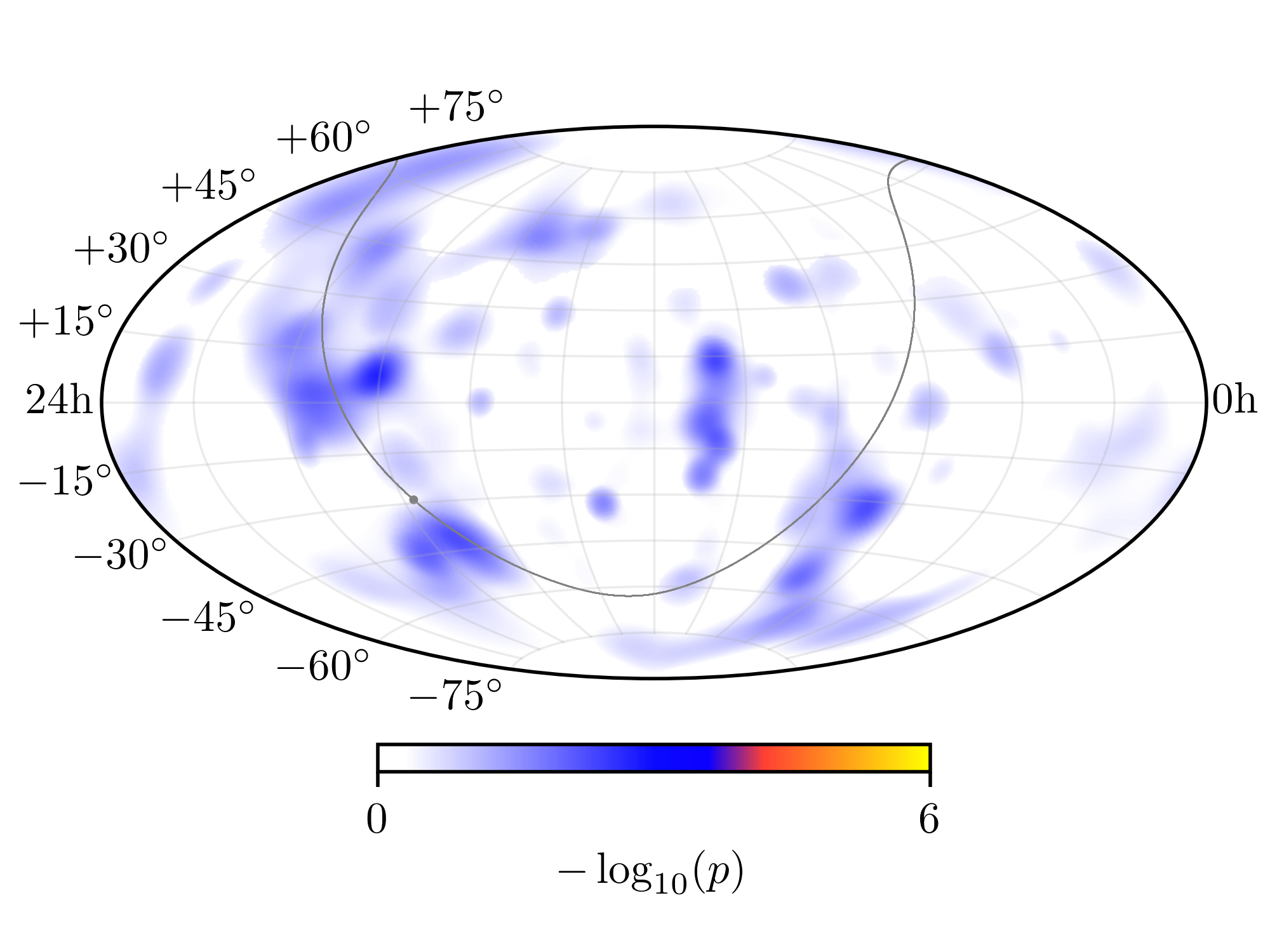}
  \caption{
    Pre-trial significance as a function of direction, in equatorial
    coordinates (J2000), for the all-sky scan.  The galactic plane
    (center) is indicated by a grey curve (dot).
  }
  \label{fig:results:allsky}
\end{figure}

The result of the unbiased all-sky scan is shown in
Figure~\ref{fig:results:allsky}.  The most significant source candidate was
found at $(\alpha,\delta)=(271.23^\circ,7.78^\circ)$ with a pre-trial p-value
of $1.8\times10^{-3}$ ($2.9\sigma$), corresponding to a post-trial p-value of
0.69.

The results of the source candidate catalog scan are tabulated
Table~\ref{tab:cat}.  The most significant source was RX~J1713.7-3946, a
well-known SNR that is also included in the SNR-alone catalog.  For this source
candidate we found a pre-trial p-value of $5.0\times10^{-3}$ ($2.6\sigma$),
corresponding to a post-trial p-value of 0.28.  Flux upper limits for each
source are plotted, along with the sensitivity and $5\sigma$ discovery
potential of this analysis, in Figure~\ref{fig:results:sensdisclim} as a
function of source declination for each of the benchmark point source spectra
discussed in the previous section.  For the one monitored source, \TXS, we find
$n_s=0$.  Note that the measured flux for \TXS\ is just $E^2\cdot dN/dE \sim
\unit[10^{-12}]{TeV\,cm^{-2}\,s^{-1}}$, or about $5\times$ lower than the
cascade sensitivity at $\delta=5.69^\circ$, and thus the null result we find
here is consistent with previous results~\citep{IceCube:2018cha}.

\begin{figure}[t]
  \centering
  \includegraphics[width=\columnwidth]{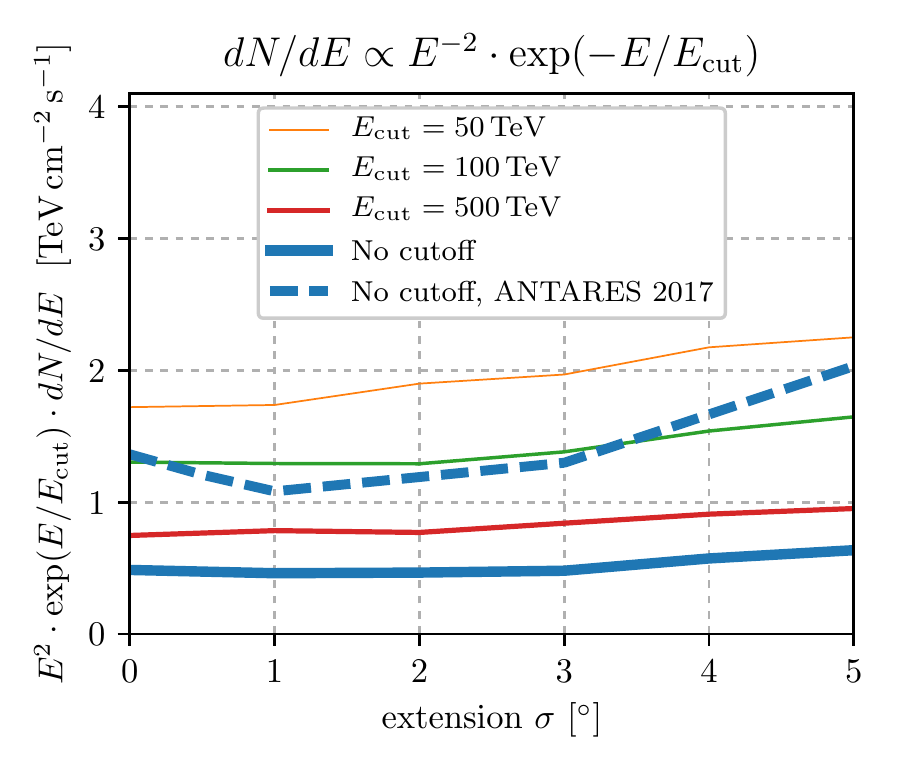}
  \caption{
    Per-flavor upper limit for Sagitarius A$^*$, as a function of possible
    angular extension, including for some choices of a possible exponential
    cutoff energy, $E_\text{cut}$.  ANTARES curves are taken from
    \cite{Albert:2017ohr}.
  }
  \label{fig:concout:sgrA}
\end{figure}

We set constraints on extended emission in the vicinity of the supermassive
black hole at the center of the Galaxy, Sagitarius A$^*$, in
Figure~\ref{fig:concout:sgrA}.  For this object we find a small but non-zero
best fit ($p_\text{pre}=0.357$).  We then compute ULs, assuming a spectrum of
the form $dN/dE\propto E^{-2}\cdot\exp(E/E_\text{cut})$ for various choices of
$E_\text{cut}$, as a function of possible Gaussian source extension,
$\sigma_{\text{Sgr~A}^*}\in[0,5^\circ]$.  In these calculations, we include the
source extension only in the signal simulation but not in the likelihood test.
The relative independence of this result with respect to assumed source
extension underscores the importance of atmospheric background rejection at the
event selection level, relative to per-event angular reconstruction, in the
overall performance of this analysis.

The results of the SNR stacking analyses are shown in
Table~\ref{tab:sensres:snr}.  We find $n_s=0$ for SNR with PWN and mild
excesses for the other two catalogs, the most significant of which is an excess
with $p=0.082$ for SNR alone.  The SNR-with-PWN category is the only one for
which this analysis finds a sensitivity flux below the previous UL from the
track analysis~\citep{Aartsen:2017ujz}; the UL found here constitutes a
reduction of $\sim50\%$.

The results of the diffuse Galactic extended template analyses are shown in
Table~\ref{tab:sensres:gp}.  The primary hypothesis test, for emission
following the \KRAg\ model, was also the most significant with a p-value of
0.021 ($2.0\sigma$) and a best-fit flux\footnote{Note that fitted fluxes,
  unlike ULs, are central values and are thus not subject to the penalty
factors described in \ref{sec:methsens:sys}} of $0.85\times\KRAg$.  The
best-fit fluxes for each template are consistent with ULs set by previous work
\citep{Aartsen:2017ujz,Albert:2018vxw}.

\begin{figure}[b]
  \centering
  \includegraphics[width=\columnwidth]{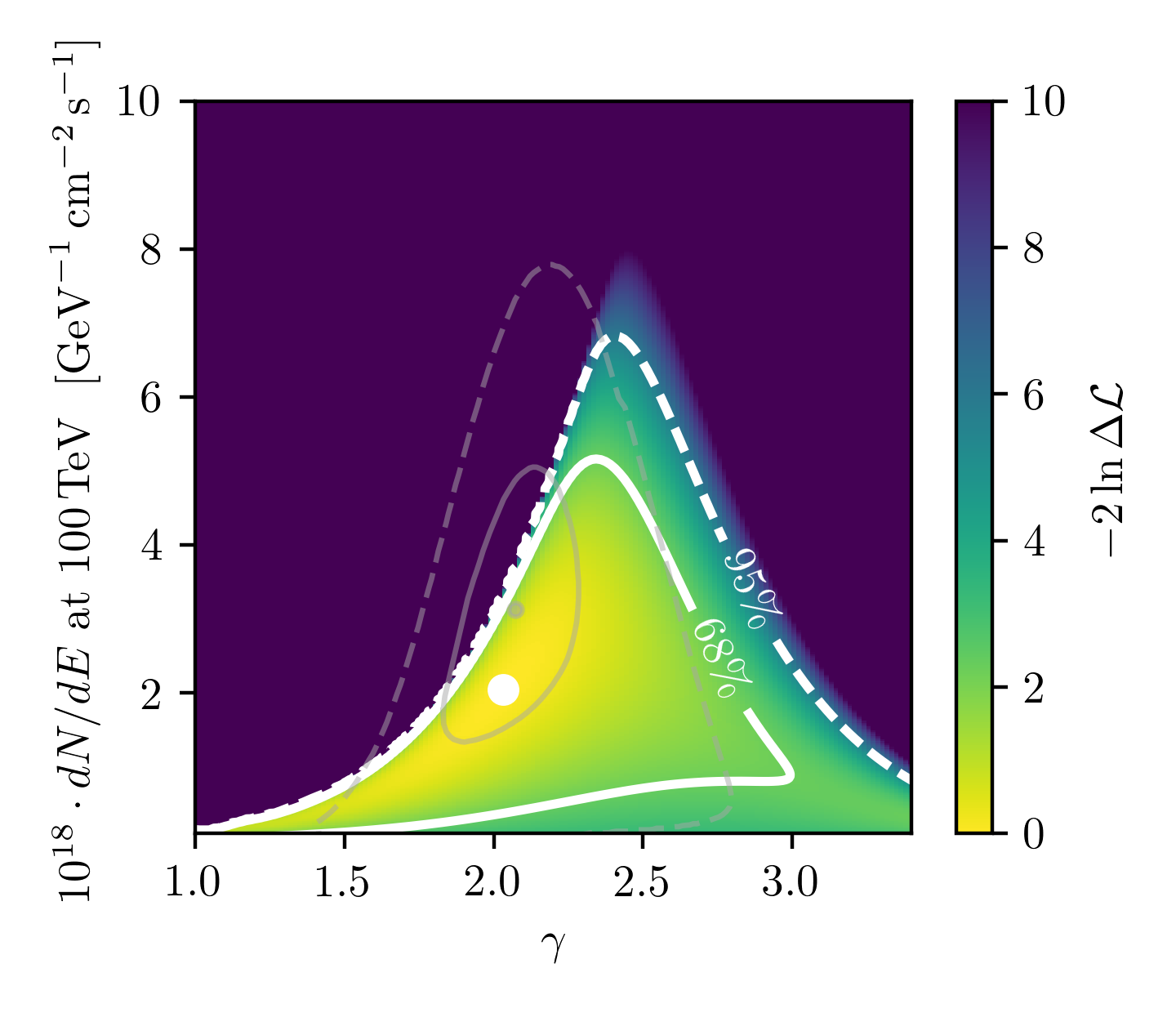}
  \caption{
    \emph{A posteriori} likelihood scan of spatially-integrated, per-flavor
    Galactic flux as a function of normalization and spectral index.  Solid
    (dashed) contours indicate 68\% (95\%) confidence regions.  Grey
    contours show the result of past IceCube work using tracks from the
    northern sky \citep{Aartsen:2017ujz}, for comparison.
  }
  \label{fig:concout:gp}
\end{figure}

Prior to this analysis, the most significant ($1.5\sigma$) indication for
diffuse Galactic emission came from an IceCube analysis using a
spatially-binned method and only events originating in the northern sky in
order to constrain the spectrum of possible emission following the \FermiL\
$\pi^0$ template~\citep{Aartsen:2017ujz}.  As an \emph{a posteriori} test, we
extend the template analysis described in Section~\ref{sec:methsens:meth} to
include the spectral index $\gamma$ as a free parameter.  A 2D scan of the
resulting likelihood for the \FermiL\ $\pi^0$ model is shown in
Figure~\ref{fig:concout:gp}, with contours from the spatially-binned track
analysis shown for comparison.  In both analyses, the best fit is obtained for
a harder spectrum close to $\gamma=2$, with both normalization and spectral
index consistent within less than $1\sigma$.  These independent results would
remain statistically insignificant even under a combined analysis.
Nevertheless, they are consistent with each other and with a possible
astrophysical signal, potentially imperfectly tracing the spatial dependence
prescribed by the \KRAga\ and \FermiL\ $\pi^0$ models, at a level only starting
to approach the reach of existing detectors and methods.

\begin{figure}[ht]
  \centering
  \includegraphics[width=\columnwidth]{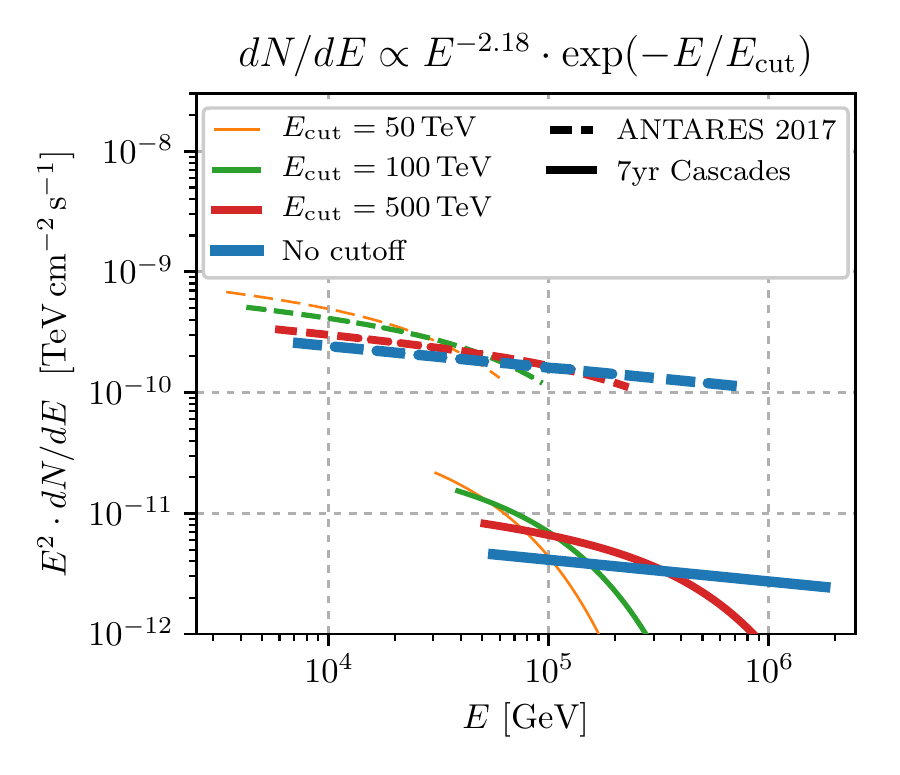}
  \caption{
    Per-flavor upper limits, shown as functions of neutrino energy, for
    emission from the \Fermi\ Bubbles.  Various exponential cutoffs are
    considered as indicated in the legend.  The horizontal span of each curve
    indicates the energy range containing 90\% of signal events for each
    spectral hypothesis based on signal MC.  Space-integrated fluxes are shown;
    our \Fermi\ bubble template spans a total solid angle of $\unit[1.18]{sr}$
    while the template used by ANTARES \citep{Hallmann:2017mac} spans a total solid
    angle of $\sim\unit[0.66]{sr}$.
  }
  \label{fig:concout:fb}
\end{figure}

For emission from the \Fermi\ bubbles, we obtain $n_s=5.2$, with a p-value of
$0.30~(0.51\sigma)$.  Flux upper limits based on these tests are shown in
Figure~\ref{fig:concout:fb}.  In the absence of significant emission, we set
the most stringent limits to date on possible high energy neutrino emission
from this intriguing structure.

\section{Conclusion and Outlook}
\label{sec:concout}

In this work, we apply a novel NN reconstruction to seven years of IceCube
cascade data in order to search for high energy neutrino emission from a number
of astrophysical source candidates.  By improving the angular resolution and
time-integrated signal acceptance with respect to our previous analysis using
two years of data~\citep{Aartsen:2017eiu}, we obtain significant gains in
sensitivity, with the best sensitivity of any experiment to date for sources
concentrated in the southern sky.  Nevertheless, we did not find significant
evidence for emission from any of the sources considered.

While we have considered several neutrino source candidates, the ensemble of
tests is far from exhaustive.  We have begun to revisit multi-wavelength EM
data in an effort to identify new catalogs of sources of interest for
individual and stacking analyses.  Furthermore, as in our previous paper
\citep{Aartsen:2017eiu}, we have still used IceCube cascades primarily in just
time-integrated analyses.  In future work we intend to explore time-dependent
source candidates, including e.g.~high-variability blazars as well as
transients such as gravitational wave candidates reported by Advanced LIGO
\citep{Martynov:2016fzi}.  The NN reconstruction is especially promising for
rapid follow-up of transient source candidates because once the NN is trained,
compute time for the reconstruction is negligible.

In future work, we plan to revisit the event selection criteria.  The selection
used in this paper already achieves very good rejection of atmospheric
backgrounds using explicit cuts on low-level parameters in the data.  However,
it is possible to improve the signal acceptance by including machine learning
methods not only in the cascade reconstruction but in the event selection as
well~\citep[e.g.][]{Niederhausen:2017mjk}.

Finally, we have deliberately attempted to maintain statistical independence
between this analysis and others performed using IceCube tracks.  We have
separately developed multiple throughgoing
\citep[e.g.][]{Aartsen:2016oji,Aartsen:2016xlq} and starting
\citep{Aartsen:2016tpb,Aartsen:2019ono} track selections, each with differing
energy- and declination-dependent background rates and signal acceptances.
Combined analyses using tracks and cascades may offer the best sensitivity
achievable using the existing IceCube detector alone.  Joint IceCube--ANTARES
analyses so far have not included IceCube cascades
(\citet{Adrian-Martinez:2015ver}, updated results in preparation).  All-flavor,
multi-detector analysis will likely give the best possible sensitivity in a
future analysis.


\begin{acknowledgements}
\emph{Acknowledgements:}
The IceCube collaboration acknowledges the significant contributions to this manu\-script
from Michael Richman.  The authors gratefully acknowledge the support from the following
agencies and institutions:
USA {\textendash} U.S. National Science Foundation-Office of
Polar Programs, U.S. National Science Foundation-Physics Division, Wisconsin Alumni
Research Foundation, Center for High Throughput Computing (CHTC) at the University of
Wisconsin-Madison, Open Science Grid (OSG), Extreme Science and Engineering Discovery
Environment (XSEDE), U.S. Department of Energy-National Energy Research Scientific
Computing Center, Particle astrophysics research computing center at the University of
Maryland, Institute for Cyber-Enabled Research at Michigan State University, and
Astroparticle physics computational facility at Marquette University; Belgium
{\textendash} Funds for Scientific Research (FRS-FNRS and FWO), FWO Odysseus and Big
Science programmes, and Belgian Federal Science Policy Office (Belspo); Germany
{\textendash} Bundesministerium f{\"u}r Bildung und Forschung (BMBF), Deutsche
Forschungsgemeinschaft (DFG), Helmholtz Alliance for Astroparticle Physics (HAP),
Initiative and Networking Fund of the Helmholtz Association, Deutsches Elektronen
Synchrotron (DESY), and High Performance Computing cluster of the RWTH Aachen; Sweden
{\textendash} Swedish Research Council, Swedish Polar Research Secretariat, Swedish
National Infrastructure for Computing (SNIC), and Knut and Alice Wallenberg Foundation;
Australia {\textendash} Australian Research Council; Canada {\textendash} Natural Sciences
and Engineering Research Council of Canada, Calcul Qu{\'e}bec, Compute Ontario, Canada
Foundation for Innovation, WestGrid, and Compute Canada; Denmark {\textendash} Villum
Fonden, Danish National Research Foundation (DNRF), Carlsberg Foundation; New Zealand
{\textendash} Marsden Fund; Japan {\textendash} Japan Society for Promotion of Science
(JSPS) and Institute for Global Prominent Research (IGPR) of Chiba University; Korea
{\textendash} National Research Foundation of Korea (NRF); Switzerland {\textendash} Swiss
National Science Foundation (SNSF).
\end{acknowledgements}



\clearpage

\begin{center}
  \startlongtable
  \begin{deluxetable*}{llrrrrr rrr}
    \vspace{2em}
    \tablecaption{Summary of the source catalog search.
      The type, common name, and equatorial coordinates
      (J2000) are shown for each object.  Where non-null ($\hat n_s>0$)
      results are found, the pre-trials significance $p_\text{pre}$ and
      best-fit $\hat n_s$ and $\hat\gamma$ are given.  ULs are expressed as
      $E^2\cdot dN/dE$, in units $\unit[10^{-12}]{TeV}$, at
      $E=\unit[100]{TeV}$ for unbroken $E^{-2}$ and $E^{-3}$ spectra
      ($\Phi_2$ and $\Phi_3$ respectively) as well as at
      $E\ll\unit[100]{TeV}$ for a spectrum with $dN/dE\propto
      E^{-2}\cdot\exp(E/\unit[100]{TeV})$ ($\Phi_{2C}$).
    \vspace{.2em}\label{tab:cat}}
    \tablewidth{0pt}
    \tablehead{
      Type & Source & $\alpha~(^\circ)$ & $\delta~(^\circ)$
      & $p_\text{pre}$ & $\hat n_s$ & $\hat\gamma$
      & $\Phi_2$ & $\Phi_3$ & $\Phi_{2C}$
    }
    \startdata
\hline
BL Lac                     &PKS 2005-489         &     302.37 & $    -48.82$ & 0.222    &    7.0 &     3.8 & 5.3 & 4.1 & 15 \\
                           &PKS 0537-441         &      84.71 & $    -44.09$ & $\cdot\cdot\cdot$ &    0.0 & $\cdot\cdot\cdot$ & 3.6 & 2.6 & 10 \\
                           &PKS 0426-380         &      67.17 & $    -37.93$ & $\cdot\cdot\cdot$ &    0.0 & $\cdot\cdot\cdot$ & 3.6 & 2.7 & 10 \\
                           &PKS 0548-322         &      87.67 & $    -32.27$ & 0.457    &    0.5 &     2.4 & 4.1 & 3.1 & 11 \\
                           &H 2356-309           &     359.78 & $    -30.63$ & 0.452    &    0.0 & $\cdot\cdot\cdot$ & 3.8 & 2.8 & 11 \\
                           &PKS 2155-304         &     329.72 & $    -30.22$ & 0.452    &    0.0 & $\cdot\cdot\cdot$ & 3.8 & 2.8 & 10 \\
                           &1ES 1101-232         &     165.91 & $    -23.49$ & 0.030    &    3.6 &     2.3 & 9.2 & 7.4 & 25 \\
                           &1ES 0347-121         &      57.35 & $    -11.99$ & $\cdot\cdot\cdot$ &    0.0 & $\cdot\cdot\cdot$ & 3.8 & 3.3 & 10 \\
                           &PKS 0235+164         &      39.66 & $     16.62$ & $\cdot\cdot\cdot$ &    0.4 &     3.3 & 5.6 & 3.6 & 11 \\
                           &1ES 0229+200         &      38.20 & $     20.29$ & 0.459    &    0.0 & $\cdot\cdot\cdot$ & 5.8 & 3.7 & 12 \\
                           &W Comae              &     185.38 & $     28.23$ & 0.475    &    0.0 & $\cdot\cdot\cdot$ & 6.0 & 3.4 & 11 \\
                           &Mrk 421              &     166.11 & $     38.21$ & 0.373    &    0.0 & $\cdot\cdot\cdot$ & 7.0 & 3.5 & 13 \\
                           &Mrk 501              &     253.47 & $     39.76$ & 0.373    &    0.0 & $\cdot\cdot\cdot$ & 7.1 & 3.4 & 13 \\
                           &BL Lac               &     330.68 & $     42.28$ & 0.160    &    6.5 &     3.4 & 9.9 & 5.0 & 18 \\
                           &H 1426+428           &     217.14 & $     42.67$ & 0.311    &    1.1 &     2.8 & 7.9 & 3.8 & 14 \\
                           &3C66A                &      35.67 & $     43.04$ & 0.351    &    0.0 & $\cdot\cdot\cdot$ & 7.4 & 3.5 & 13 \\
                           &1ES 2344+514         &     356.77 & $     51.70$ & 0.119    &    7.5 &     4.0 & 13 & 5.5 & 23 \\
                           &1ES 1959+650         &     300.00 & $     65.15$ & 0.137    &    6.1 &     4.0 & 20 & 5.2 & 30 \\
                           &S5 0716+71           &     110.47 & $     71.34$ & 0.480    &    1.5 &     3.3 & 13 & 2.9 & 20 \\
\hline
Flat Spectrum Radio Quasar &PKS 1454-354         &     224.36 & $    -35.65$ & 0.487    &    0.6 &     3.4 & 3.6 & 2.8 & 10 \\
                           &PKS 1622-297         &     246.52 & $    -29.86$ & 0.315    &    4.2 &     4.0 & 4.8 & 3.7 & 13 \\
                           &PKS 0454-234         &      74.27 & $    -23.43$ & 0.483    &    0.0 & $\cdot\cdot\cdot$ & 3.6 & 2.9 & 9.9 \\
                           &QSO 1730-130         &     263.26 & $    -13.08$ & 0.162    &    1.2 &     1.7 & 6.5 & 5.9 & 19 \\
                           &PKS 0727-11          &     112.58 & $    -11.70$ & 0.293    &   11.1 &     3.6 & 5.5 & 4.8 & 15 \\
                           &PKS 1406-076         &     212.23 & $     -7.87$ & $\cdot\cdot\cdot$ &    0.0 & $\cdot\cdot\cdot$ & 3.8 & 3.4 & 10 \\
                           &QSO 2022-077         &     306.42 & $     -7.64$ & $\cdot\cdot\cdot$ &    0.0 & $\cdot\cdot\cdot$ & 3.8 & 3.3 & 10 \\
                           &3C279                &     194.05 & $     -5.79$ & $\cdot\cdot\cdot$ &    1.1 &     2.5 & 3.9 & 3.4 & 10 \\
                           &3C 273               &     187.28 & $      2.05$ & 0.435    &    2.3 &     2.5 & 4.6 & 3.9 & 11 \\
                           &PKS 1502+106         &     226.10 & $     10.49$ & $\cdot\cdot\cdot$ &    2.7 &     3.8 & 5.3 & 3.7 & 11 \\
                           &PKS 0528+134         &      82.73 & $     13.53$ & $\cdot\cdot\cdot$ &    0.0 & $\cdot\cdot\cdot$ & 5.4 & 3.7 & 12 \\
                           &3C 454.3             &     343.49 & $     16.15$ & 0.288    &    1.9 &     2.1 & 8.0 & 5.3 & 17 \\
                           &4C 38.41             &     248.81 & $     38.13$ & 0.373    &    0.0 & $\cdot\cdot\cdot$ & 7.1 & 3.5 & 13 \\
\hline
Galactic Center            &Sgr A*               &     266.42 & $    -29.01$ & 0.357    &    2.2 &     3.0 & 4.5 & 3.5 & 12 \\
\hline
HMXB/mqso                  &Cir X-1              &     230.17 & $    -57.17$ & 0.400    &    0.0 & $\cdot\cdot\cdot$ & 3.7 & 2.5 & 11 \\
                           &GX 339-4             &     255.70 & $    -48.79$ & 0.016    &    5.9 &     2.1 & 9.2 & 6.6 & 26 \\
                           &LS 5039              &     276.56 & $    -14.83$ & 0.459    &    4.6 &     3.6 & 4.3 & 3.4 & 11 \\
                           &SS433                &     287.96 & $      4.98$ & 0.011    &   30.9 &     3.1 & 14 & 10.0 & 33 \\
                           &HESS J0632+057       &      98.25 & $      5.80$ & $\cdot\cdot\cdot$ &    0.0 & $\cdot\cdot\cdot$ & 4.7 & 3.4 & 11 \\
                           &Cyg X-1              &     299.59 & $     35.20$ & 0.130    &    8.6 &     3.0 & 11 & 5.4 & 20 \\
                           &Cyg X-3              &     308.11 & $     40.96$ & 0.150    &    7.7 &     3.2 & 11 & 5.0 & 19 \\
                           &LSI 303              &      40.13 & $     61.23$ & $\cdot\cdot\cdot$ &    0.0 & $\cdot\cdot\cdot$ & 10 & 2.9 & 16 \\
\hline
Massive Star Cluster       &HESS J1614-518       &      63.58 & $    -51.82$ & $\cdot\cdot\cdot$ &    0.0 & $\cdot\cdot\cdot$ & 3.6 & 2.5 & 11 \\
\hline
Not Identified             &HESS J1507-622       &     226.72 & $    -62.34$ & 0.287    &    0.0 & $\cdot\cdot\cdot$ & 4.1 & 2.8 & 12 \\
                           &HESS J1503-582       &     226.46 & $    -58.74$ & 0.353    &    0.0 & $\cdot\cdot\cdot$ & 3.9 & 2.7 & 11 \\
                           &HESS J1741-302       &     265.25 & $    -30.20$ & 0.201    &    5.5 &     3.0 & 5.8 & 4.4 & 16 \\
                           &HESS J1837-069       &      98.69 & $     -8.76$ & 0.470    &    4.3 &     3.4 & 3.9 & 3.5 & 10 \\
                           &HESS J1834-087       &     278.69 & $     -8.76$ & 0.102    &   22.3 &     3.5 & 7.5 & 6.6 & 20 \\
                           &MGRO J1908+06        &     286.98 & $      6.27$ & 0.018    &   28.3 &     3.0 & 14 & 9.6 & 32 \\
\hline
Pulsar Wind Nebula         &HESS J1356-645       &     209.00 & $    -64.50$ & 0.286    &    0.0 & $\cdot\cdot\cdot$ & 3.8 & 2.8 & 12 \\
                           &PSR B1259-63         &     197.55 & $    -63.52$ & 0.287    &    0.0 & $\cdot\cdot\cdot$ & 4.0 & 2.8 & 12 \\
                           &HESS J1303-631       &     195.74 & $    -63.20$ & 0.287    &    0.0 & $\cdot\cdot\cdot$ & 4.0 & 2.8 & 12 \\
                           &MSH 15-52            &     228.53 & $    -59.16$ & 0.353    &    0.0 & $\cdot\cdot\cdot$ & 3.9 & 2.7 & 11 \\
                           &HESS J1023-575       &     155.83 & $    -57.76$ & 0.096    &    4.7 &     4.0 & 5.7 & 4.4 & 17 \\
                           &HESS J1616-508       &     243.78 & $    -51.40$ & 0.146    &    1.7 &     1.7 & 6.1 & 4.4 & 18 \\
                           &HESS J1632-478       &     248.04 & $    -47.82$ & 0.044    &    3.8 &     2.0 & 8.3 & 6.0 & 24 \\
                           &Vela X               &     128.75 & $    -45.60$ & $\cdot\cdot\cdot$ &    0.1 &     2.0 & 3.8 & 2.7 & 11 \\
                           &Geminga              &      98.48 & $     17.77$ & $\cdot\cdot\cdot$ &    0.0 & $\cdot\cdot\cdot$ & 5.5 & 3.7 & 11 \\
                           &Crab Nebula          &      83.63 & $     22.01$ & 0.461    &    0.0 & $\cdot\cdot\cdot$ & 6.0 & 3.6 & 12 \\
                           &MGRO J2019+37        &     305.22 & $     36.83$ & 0.182    &    6.8 &     3.0 & 9.8 & 4.9 & 18 \\
\hline
Seyfert Galaxy             &ESO 139-G12          &     264.41 & $    -59.94$ & 0.247    &    1.6 &     2.6 & 4.6 & 3.3 & 13 \\
\hline
Star Formation Region      &Cyg OB2              &     308.08 & $     41.51$ & 0.144    &    8.0 &     3.2 & 11 & 5.0 & 19 \\
\hline
Starburst/Radio Galaxy     &Cen A                &     201.36 & $    -43.02$ & $\cdot\cdot\cdot$ &    0.0 & $\cdot\cdot\cdot$ & 3.7 & 2.7 & 10 \\
                           &M87                  &     187.71 & $     12.39$ & 0.305    &    3.2 &     2.4 & 7.6 & 5.2 & 17 \\
                           &3C 123.0             &      69.27 & $     29.67$ & 0.302    &    1.0 &     2.2 & 8.0 & 4.7 & 16 \\
                           &Cyg A                &     299.87 & $     40.73$ & 0.050    &   11.2 &     3.1 & 13 & 6.4 & 24 \\
                           &NGC 1275             &      49.95 & $     41.51$ & 0.361    &    0.0 & $\cdot\cdot\cdot$ & 7.6 & 3.5 & 13 \\
                           &M82                  &     148.97 & $     69.68$ & 0.265    &    3.4 &     3.2 & 19 & 4.2 & 28 \\
\hline
Supernova Remnant          &RCW 86               &     220.68 & $    -62.48$ & 0.287    &    0.0 & $\cdot\cdot\cdot$ & 4.1 & 2.8 & 12 \\
                           &RX J0852.0-4622      &     133.00 & $    -46.37$ & $\cdot\cdot\cdot$ &    0.0 & $\cdot\cdot\cdot$ & 3.7 & 2.5 & 11 \\
                           &\tablenotemark{$\dagger$}RX J1713.7-3946      &     258.25 & $    -39.75$ & 0.005    &   10.8 &     2.5 & 11 & 8.6 & 32 \\
                           &W28                  &     270.43 & $    -23.34$ & 0.238    &    0.8 &     1.6 & 5.6 & 4.7 & 16 \\
                           &IC443                &      94.18 & $     22.53$ & 0.461    &    0.0 & $\cdot\cdot\cdot$ & 6.1 & 3.7 & 12 \\
                           &Cas A                &     350.85 & $     58.81$ & 0.028    &   12.4 &     4.0 & 24 & 7.0 & 38 \\
                           &TYCHO                &       6.36 & $     64.18$ & 0.069    &    9.5 &     3.7 & 22 & 6.0 & 34 \\
    \enddata
    \tablenotetext{\dagger}{Most significant source in the catalog, yielding
    $p_\text{post}=0.28$.}
  \end{deluxetable*}
\end{center}

\bibliographystyle{apj}
\bibliography{bib}


\end{document}